\documentclass[%
 reprint,
 amsmath,amssymb,
 aps
]{revtex4-2}

\usepackage{graphicx}
\usepackage{dcolumn}
\usepackage{bm}
\usepackage{color}
\usepackage{cancel}
\usepackage{comment}

\usepackage[
  colorlinks=true,
  linkcolor=blue,
  citecolor=blue,
  urlcolor=blue
]{hyperref}

\DeclareRobustCommand{\erase}{\bgroup\markoverwith{\textcolor{red}{\rule[.5ex]{2pt}{0.4pt}}}\ULon}

\graphicspath{ {Pic/} }

\begin{document}

\newtheorem{definition}{Definition}[section]
\newcommand{\be}{\begin{equation}}
\newcommand{\ee}{\end{equation}}
\newcommand{\bea}{\begin{eqnarray}}
\newcommand{\eea}{\end{eqnarray}}
\newcommand{\LE}{\left[}
\newcommand{\R}{\right]}
\newcommand{\nn}{\nonumber}
\newcommand{\Tr}{\text{Tr}}
\newcommand{\N}{\mathcal{N}}
\newcommand{\G}{\Gamma}
\newcommand{\vf}{\varphi}
\newcommand{\LL}{\mathcal{L}}
\newcommand{\Op}{\mathcal{O}}
\newcommand{\HH}{\mathcal{H}}
\newcommand{\arctanh}{\text{arctanh}}
\newcommand{\up}{\uparrow}
\newcommand{\down}{\downarrow}
\newcommand{\ket}[1]{\left| #1 \right>}
\newcommand{\bra}[1]{\left< #1 \right|}
\newcommand{\ketbra}[1]{\left|#1\right>\left<#1\right|}
\newcommand{\rd}{\partial}
\newcommand{\del}{\partial}
\newcommand{\ba}{\begin{eqnarray}}
\newcommand{\ea}{\end{eqnarray}}
\newcommand{\db}{\bar{\partial}}
\newcommand{\we}{\wedge}
\newcommand{\ca}{\mathcal}
\newcommand{\lr}{\leftrightarrow}
\newcommand{\f}{\frac}
\newcommand{\s}{\sqrt}
\newcommand{\vp}{\varphi}
\newcommand{\hvp}{\hat{\varphi}}
\newcommand{\tvp}{\tilde{\varphi}}
\newcommand{\tp}{\tilde{\phi}}
\newcommand{\ti}{\tilde}
\newcommand{\pr}{\propto}
\newcommand{\mb}{\mathbf}
\newcommand{\ddd}{\cdot\cdot\cdot}
\newcommand{\no}{\nonumber \\}
\newcommand{\la}{\langle}
\newcommand{\lb}{\rangle}
\newcommand{\ep}{\epsilon}
\newcommand{\rmi}{\text{i}}
\newcommand{\rme}{\text{e}}
\newcommand{\rmd}{\text{d}}
 \def\we{\wedge}
 \def\lr{\leftrightarrow}
 \def\f {\frac}
 \def\ti{\tilde}
 \def\ap{\alpha}
 \def\pr{\propto}
 \def\mb{\mathbf}
 \def\ddd{\cdot\cdot\cdot}
 \def\no{\nonumber \\}
 \def\la{\langle}
 \def\lb{\rangle}
 \def\ep{\epsilon}
\newcommand{\mcl}{\mathcal}
 \def\g{\gamma}
\def\Tr{\text{tr}}

\newcommand{\RY}[1]{{\color{red} \textbf{RY}:~#1}}
\newcommand{\KN}[1]{{\color{blue} \textbf{KN}:~#1}}


\title{
Inhomogeneous entanglement structure in \\ monoaxial chiral ferromagnetic quantum spin chain
}

\author{Kentaro Nishimura$^{1,2,3}$, Ryosuke Yoshii$^{4,2,3}$}
\affiliation{
$^1$Institute of Science and Technology, Niigata University, Niigata 950-2181, Japan \\
$^2$Research and Education Center for Natural Sciences, Keio University, 4-1-1 Hiyoshi, Yokohama, Kanagawa 223-8521, Japan \\
$^3$International Institute for Sustainability with Knotted Chiral Meta Matter (SKCM$^2$), Hiroshima University, 1-3-2 Kagamiyama, Higashi-Hiroshima, Hiroshima 739-8511, Japan \\
$^4$Center for Liberal Arts and Sciences, Sanyo-Onoda City University, 1-1-1 Daigaku-Dori, Sanyo-Onoda, Yamaguchi 756-0884, Japan
}

\begin{abstract}
Chiral magnets, characterized by inhomogeneous magnetic moment arrangements, have attracted significant attention recently due to their topological orders, such as magnetic skyrmion lattices and chiral soliton lattices.
In this work, we investigate the entanglement entropy of \textit{quantum} chiral magnets and demonstrate that it reflects the inhomogeneous nature of the ground state.
We perform numerical simulations of a one-dimensional monoaxial chiral ferromagnetic chain with Zeeman term using the density matrix renormalization group method.
Our results show that the entanglement entropy exhibits oscillatory behavior, which can be tuned by varying the external magnetic field. Analysis of the local magnetization and spin chirality further confirms that these oscillations correspond to solitonic structures. 
Moreover, our findings suggest that the entanglement entropy can serve as a probe for detecting the vacuum structure, providing new insights into quantum correlations.
\end{abstract}

\maketitle

\section{Introduction} 
Topological solitons are spatially localized field configurations characterized by nontrivial topological charges.
The well-known examples include sine-Gordon solitons, baby Skyrmions, vortices, and instantons.
These solitons are ubiquitous objects that arise not only in high-energy physics---such as cosmology, particle physics, and nuclear physics---but also in condensed matter systems, including spin systems and superfluids.
It is widely recognized that topological solitons play a crucial role in understanding various physical phenomena (see famous textbooks \cite{Vilenkin:2000jqa,Manton:2004tk,Weinberg:2012pjx}).

In addition to the importance of the semiclassical solitonic structure, the quantum effects of topological solitons also play an essential role in understanding a wide range of physical phenomena. 
In high-energy physics, notable examples include the Skyrmion description \cite{Witten:1983tx,Adkins:1983ya} of nucleons and the confinement mechanism \cite{Polyakov:1975rs}, while in spin systems, Ref.~\cite{PhysRevB.53.3237} provides a relevant example.
Nevertheless, the topological solitons are classical solutions to the static equations of motion, and studying their quantum aspects is not straightforward.
One of the adopted approaches is the semiclassical method, which incorporates quantum effects by quantizing fluctuations around a classical background \cite{Dashen:1974ci,Dashen:1974cj,PhysRevB.53.3237}.
However, the semiclassical approximation breaks down when quantum fluctuations become large, rendering the method inapplicable in such regimes.
Therefore, uncovering the quantum counterparts of the topological solitons and understanding their impact on the system through a fully quantum mechanical framework remains a significant and challenging problem.

From this perspective, a particularly suitable model is the monoaxial chiral magnet.
Its Hamiltonian consists of the exchange interaction, the Zeeman term, and the Dzyaloshinskii--Moriya interaction (DMI) \cite{DZYALOSHINSKY1958241,PhysRev.120.91}.
One key reason for its relevance is that, in the classical and continuum limit, the model reduces to a $(1+1)$-dimensional sine-Gordon theory with an additional total derivative term \cite{kishine2015theory}.
This theory admits soliton solutions, whose energies become negative due to the total derivative contribution. 
As a result, the ground state forms a periodic array of topological solitons, known as the chiral soliton lattice (CSL) \cite{dzyaloshinskii1965theory}, which is experimentally observed in chiral magnets \cite{PhysRevLett.108.107202}.

Interestingly, this is not merely a toy model; such an effective theory can also be found in finite-density chiral perturbation theory under strong magnetic fields \cite{Son:2007ny,Brauner:2016pko}, as well as dense $\mathrm{QCD}_2$ in the large-$N_{\mathrm{c}}$ expansion \cite{Kojo:2011fh}. 
More recently, it has been reported that, in the continuum limit, the spin system with DMI can be mapped onto a field theory in a curved background by appropriately tuning the system parameters~\cite{Kinoshita:2024ahu}.

A second reason for the suitability of this model lies in recent experimental proposals for realizing $S=1/2$ quantum spin chains with exchange interaction, DMI, and Zeeman terms using quantum simulators based on Rydberg atoms \cite{PhysRevA.110.043312}.
Therefore, investigating the quantum effects of the topological solitons in monoaxial chiral ferromagnetic spin systems not only offers the potential for experimental justification but also provides the advantage of feedback from experimental data.

The third reason is numerical feasibility:
the ground state of this system can be efficiently obtained using the density matrix renormalization group (DMRG) method \cite{PhysRevLett.69.2863,SCHOLLWOCK201196},
with the matrix product state (MPS) \cite{Perez-Garcia:2006nqo} as a variational ansatz. 
The other advantage of DMRG is its ability to keep the Hilbert space dimension small,
enabling numerical simulations with many sites.
Since the study of spatially inhomogeneous structures,
such as the topological solitons,
requires large system sizes,
this model---in which the sine-Gordon solitons naturally appear in the ground state---is particularly well-suited for such investigations.

Indeed, the chiral ferromagnetic quantum spin chains and related models have attracted significant attention in the research community. 
Notably, several studies employing fully quantum mechanical methods—such as exact diagonalization, DMRG, and analytical calculations—have reported fascinating phenomena, including the quantum many-body scars \cite{PhysRevA.110.043312} and spin parity effects \cite{PhysRevB.107.024403,PhysRevB.110.L100403}.

Since we are interested in the quantum effects on the topological solitons, we focus on the entanglement structure inside the system. Entanglement is a fundamental feature of quantum mechanics that has no counterpart in classical physics. Initially introduced to highlight the incompleteness of quantum theory—due to its nonlocal nature \cite{PhysRev.47.777} and apparent incompatibility with local realism—it is now widely regarded as a hallmark of quantum behavior \cite{PhysicsPhysiqueFizika.1.195}. 
Subsequently, this intrinsic nonlocality has been harnessed for applications in secure communication and quantum computation \cite{PhysRevLett.70.1895,Nielsen/Chuang:00}. 

More recently, entanglement entropy has emerged as a key tool for quantifying entanglement between two subsystems (see Refs.~\cite{RevModPhys.80.517,RevModPhys.81.865} and references therein). 
Its introduction has significantly advanced the study of quantum correlations across diverse systems and has found applications in various research domains (See, for instance, Refs.~\cite{LAFLORENCIE20161,RevModPhys.90.035007, PhysRevLett.90.227902}).

Entanglement entropy in a local subsystem plays a crucial role in understanding thermalization in isolated quantum systems \cite{PhysRevA.43.2046, PhysRevE.50.888}. 
One of the most widely accepted mechanisms for quantum thermalization, consistent with unitary time evolution, is the eigenstate thermalization hypothesis (ETH). 
According to ETH, while the total system evolves unitarily, the expectation value of a local observable is determined by the properties of the reduced density matrix, as the complementary degrees of freedom are traced out during measurement. 
A key characteristic of this reduced density matrix is entanglement entropy. 
Recently, the cousin of the entanglement entropy, the so-called R\'enyi entanglement entropy, has been successfully observed in Refs.\ \cite{Islam2015Measuring,11796}, and the thermalization of the subsystem is exhibited. 

Since the entanglement structure encodes information about the vacuum state of a system, its behavior has been extensively studied in various settings \cite{Calabrese:2004eu,RevModPhys.82.277}. 
However, its properties in the presence of a nontrivial background remain largely unexplored. 
In this work, we investigate a model with an inhomogeneous background and demonstrate that the entanglement entropy reflects this inhomogeneity. 
Our key findings are twofold. First, the entropy’s oscillatory pattern corresponds to soliton locations, verified by magnetization and chirality profiles. 
Second, these oscillations vanish in ferromagnetic regimes, suggesting entanglement entropy acts as a sensitive probe of vacuum inhomogeneity. 
This aligns with its role in perturbed field theories  \cite{Casini:2009sr} and offers a new lens for understanding quantum solitonic systems.
We outline the model and methods, present results, and discuss implications for quantum matter.

\section{Model and Methods}
\begin{figure}
    \centering
    \includegraphics[width=0.8\linewidth]{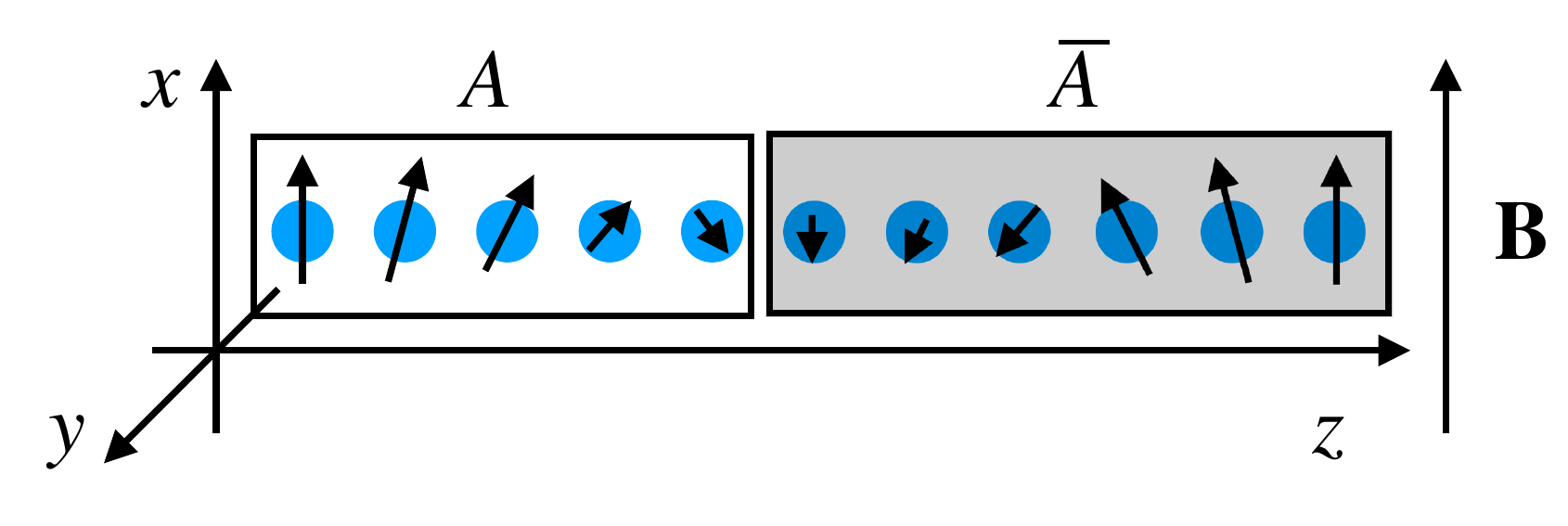}
    \caption{Schematic of the model. Sites align along the $z$-direction, with the magnetic field applied along $x$. The system is divided into subsystem $A$ (sites 1 to $i$) and its complement $\bar{A}$. }
    \label{fig:model}
\end{figure}

We consider a spin-1/2 chain along the $z$-direction with $N$ sites, governed by the following Hamiltonian: 
\begin{align}
    \hat H
    =&-\sum_{n=1}^{N-1}
    \left(
    J\delta_{ij}
    +D\epsilon_{zij}
    \right)
    \hat S_n^i  \hat S_{n+1}^j
    +B\sum_{n=1}^{N}\hat S_n^x \,,
\end{align}
where $S_n^i$ is the spin operator’s $i$-component ($i=x,y,z$), $\delta_{ij}$ and $\epsilon_{zij}$ are, respectively, the Kronecker delta and the Levi-Civita symbol, $J$ (set to $1$ as the energy scale) governs exchange interaction, $D/J=2$ is the DMI strength, and $B$ is the magnetic field along $x$. 
Here we use the Einstein summation rule. 
This model captures helical and soliton-like orders via the competition between these interactions (see Fig.\ \ref{fig:model}).

This section presents numerical results obtained using DMRG with the ITensor Library~\cite{Fishman:2020gel}. 
The simulations were performed with a maximum bond dimension of $40$ and a truncation error cutoff of $10^{-8}$ in the final sweeps.
A few hundred sweeps were typically required for convergence.
We evaluate three key observables:
\begin{itemize}
    \item{
    Entanglement entropy: $S_A = -\text{Tr} \rho_A \log \rho_A$, where $\rho_A = \text{Tr}_{\bar{A}} \rho$ is the reduced density matrix for subsystem $A$ (sites 1 to $i$).
    }
    \item{
    Magnetization: $M_n^i = \langle \hat{S}_n^i \rangle$ (total magnetization is defined as $M^i=\sum_{n=1}^{N}M^i_n$), revealing spin textures.
    }
    \item{Spin chirality: $\chi^i_n = \langle \epsilon^{ijk} \hat{S}_n^j \hat{S}_{n+1}^k \rangle$, quantifying spin twisting and soliton presence.}
\end{itemize}
Here $\langle \hat A\rangle $ stands for the expectation value of the operator $\hat A$ with respect to the ground state. 

We briefly explain the third quantity. 
In this study, the topological solitons in the chiral magnets—the structures where the spins are locally twisted due to the DM interaction—play a crucial role. 
To quantify the degree of twisting between neighboring spins, the spin chirality is an appropriate measure (see Refs.~\cite{braun1996chiral,PhysRevB.107.024403} for its role in quantum spin systems). 
In the classical limit, the spin operator $\hat{S}^i_n$ can be identified with a classical vector $S^i_n$ of fixed length $|\bm{S}_n| = 1/2$. 
Assuming the spins lie in the $xy$-plane, the $z$-component of the chirality $[\bm{S}_n \times \bm{S}_{n+1}]_z$ becomes $\frac{1}{4} \sin(\Delta \varphi_n)$, where $\Delta \varphi_n $ is the angle between neighboring spins $\Delta \varphi_n = \varphi_{n+1} - \varphi_n$. 
It reaches its maximum at $\Delta \varphi_n = \pi/2$, where the spins are maximally twisted (see appendix~\ref{app:explanation_spin_chirality} for details).

\section{Results and Discussion}
\begin{figure}
    \centering
    \includegraphics[width=1\linewidth]{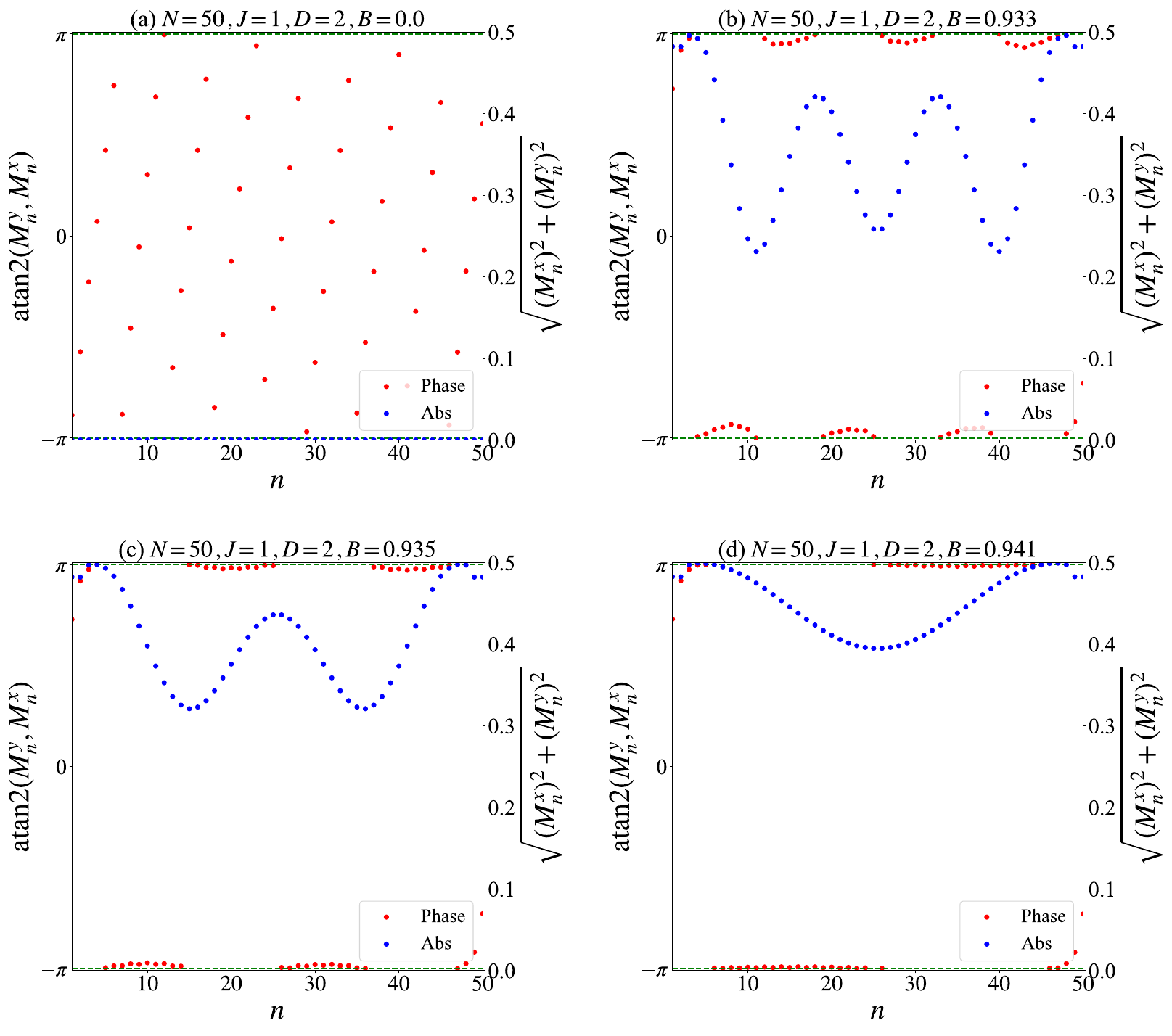}
    \caption{
    The spatial dependencies of the magnitude and the phase,
    namely $\sqrt{(M^{x}_n)^2+(M^{y}_n)^2}$ and $\mathrm{atan}2(M^y_n,M^x_n)$, are plotted in the panels for $B=0.0$ (a), $B=0.933$ (b), $B=0.935$ (c), and $B=0.941$ (d).
    }
    \label{fig:phase_abs}
\end{figure}
We simulate the model across various magnetic field strengths $B$, revealing rich entanglement and topological behavior. 
In the classical limit, the ground state at small $B$ is known to be the CSL (see \cite{kishine2015theory} and references therein). 
To assess whether this configuration persists in the quantum case,
we compute the phase $\mathrm{atan}2(M^y_n,M^x_n)$ \footnote{
The function $\operatorname{atan2}(y, x)$ returns a unique angle $\theta$ in the range $-\pi < \theta \leq \pi$ satisfying $x = r \cos \theta$ and $y = r \sin \theta$ where $r > 0$.
} and the magnitude $\sqrt{(M^{x}_n)^2+(M^{y}_n)^2}$ of the spin expectation values in the $xy$-plane.
Figure~\ref{fig:phase_abs} summarizes the results.

The phase is nearly proportional to the site index $n$ (with $\pi$ and $-\pi$ identified),
suggesting a constant pitch angle similar to the classical helical structure \cite{kishine2015theory}.
However, the magnitude remains zero across all $n$,
indicating the spin degeneracy—a clear deviation from classically expected behavior;
$\sqrt{(M^{x}_n)^2+(M^{y}_n)^2} \approx 1/2$ when $M^z_n \approx 0$ (in appendix \ref{app:local_magnetization}, we numerically show that $M^z_n\approx 0$ at all site $n$).

For $B = 0.933$, $0.935$, and $0.941$, the magnitude shows spatial modulation with 3, 2, and 1 local minima, respectively. 
While the phase suggests spins align along the $-x$ axis, this could misleadingly imply a ferromagnetic state driven by the magnetic field. 
Although the phase is a useful diagnostic in classical systems, it is much less informative in the quantum regime.

\begin{figure}
    \centering
    \includegraphics[width=1.0\linewidth]{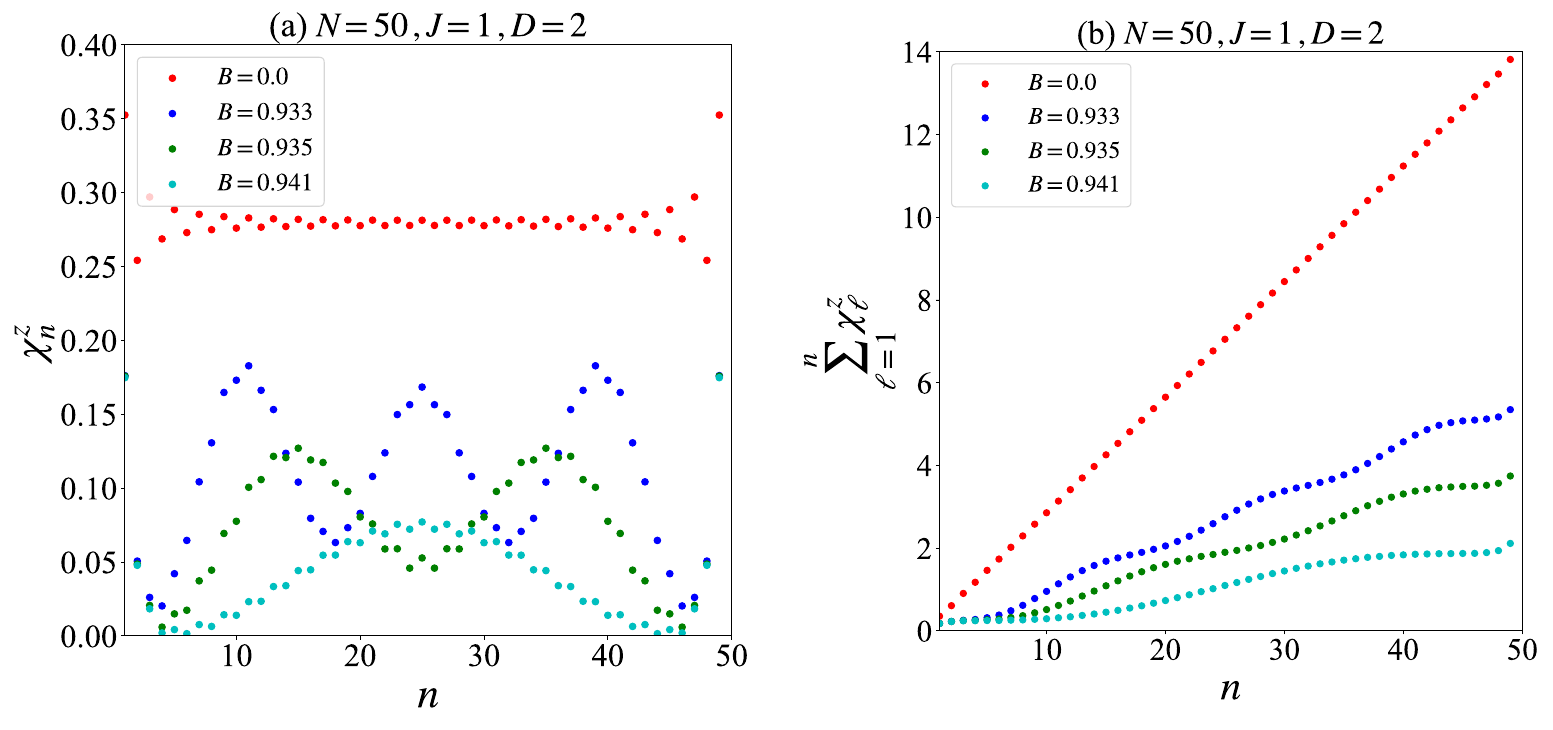}
    \caption{
    Spin chirality at site $n$ (a) and the accumulated spin chirality from the left boundary (b) for $B=0,0.933,0.935, 0.941$. 
    }
    \label{fig:spin_chirality}
\end{figure}

We examine the numerical results for the spin chirality to understand the local minima in $\sqrt{(M^{x}_n)^2+(M^{y}_n)^2}$. 
Panel~(a) in Fig.~\ref{fig:spin_chirality} shows $\chi_n^z$ (also see the numerical calculation of $\chi^{x,y}_n$ in appendix \ref{app:explanation_spin_chirality}).
For $B = 0$, it remains nearly constant except near the boundaries,
indicating a quantum spin configuration similar to the classical helical structure.

For $B = 0.933$, $0.935$, and $0.941$, $\chi_n^z$ exhibits local maxima precisely where the spin magnitude shows local minima, indicating strong twisting between neighboring spins. This behavior is reminiscent of the classical CSL state, with the peaks in $\chi_n^z$ corresponding to the quantum analogs of the classical solitons.

Panel~(b) shows the accumulated spin rotation in the $xy$-plane from site 1 to $n$. The rotation increases linearly for $B = 0$, consistent with a uniform pitch. In contrast, for the above $B$ values, sharp changes occur at the soliton positions. The number of solitons---i.e., the number of local maxima in $\chi_n^z$---determines the total rotation up to $n = 50$. 
For clarity in subsequent discussions, we define the soliton number $Q$ as the number of local maxima in $S_A$.

\begin{figure}
    \centering
    \includegraphics[width=1\linewidth]{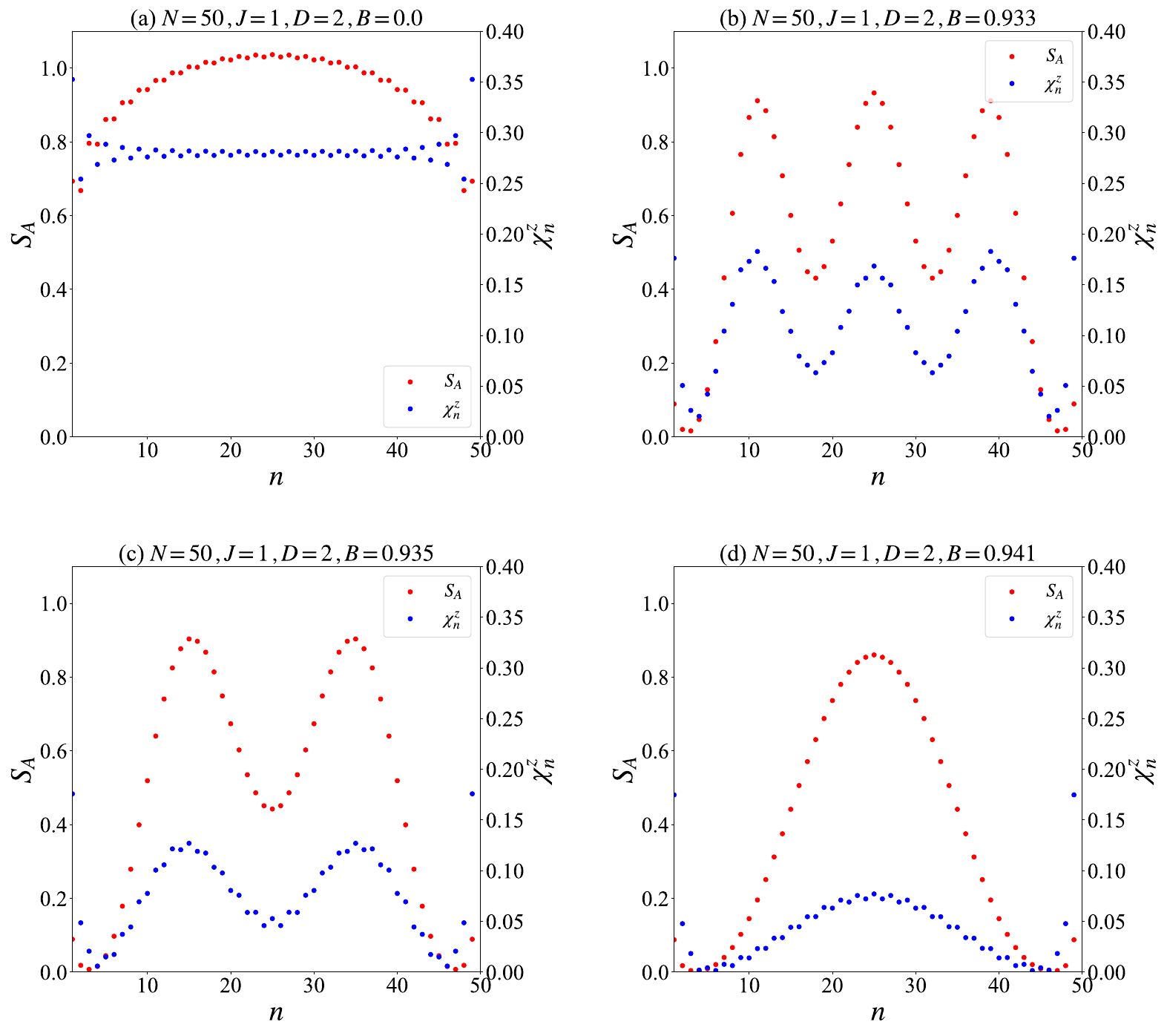}
    \caption{
    Plots of the entanglement entropy $S_A$ and $z$-component of the spin chirality $\chi^z_n$ for $N=50$ with various magnetic field strengths.
    The horizontal axis represents the subsystem size for $S_A$ and the position for $\chi^z_n$.
    Panel~(a) shows a state with constant winding, while panels~(b), (c), and (d) correspond to soliton numbers $Q$ $3$, $2$, and $1$, respectively.
    }
    \label{fig:EEandAbs}
\end{figure}

The entanglement entropy $S_A$ (Fig.\ \ref{fig:EEandAbs}) is our central result.
For $N=50$, $S_A$ versus subsystem size are plotted (Fig.\ \ref{fig:EEandAbs}(a-d)).
The peaks (dips) in $S_A$ align with soliton maxima (minima in the absolute values $\sqrt{(M^{x}_n)^2+(M^{y}_n)^2}$),
deviating from the smooth curve seen in homogeneous solutions.
At high $B$, where the spins align in a ferromagnetic manner
(
See Appendix~\ref{app:local_magnetization} for details),
$S_A$ drops to $0$, indicating that the system is described by the local product states.
We note that this oscillation is tunable: increasing $B$ reduces the soliton number $Q$, smoothing $S_A$. 
For the case of $B=0$, the solitons have huge overlapping and thus the peak-dip structure is smeared out. This causes a spatially homogeneous configuration, and thus we have a smooth curve similar to the curve typically appearing in the homogeneous case (Fig. \ref{fig:EEandAbs} (a)) \cite{PhysRevD.100.105010}.

\begin{figure}
    \centering
    \includegraphics[width=1.0\linewidth]{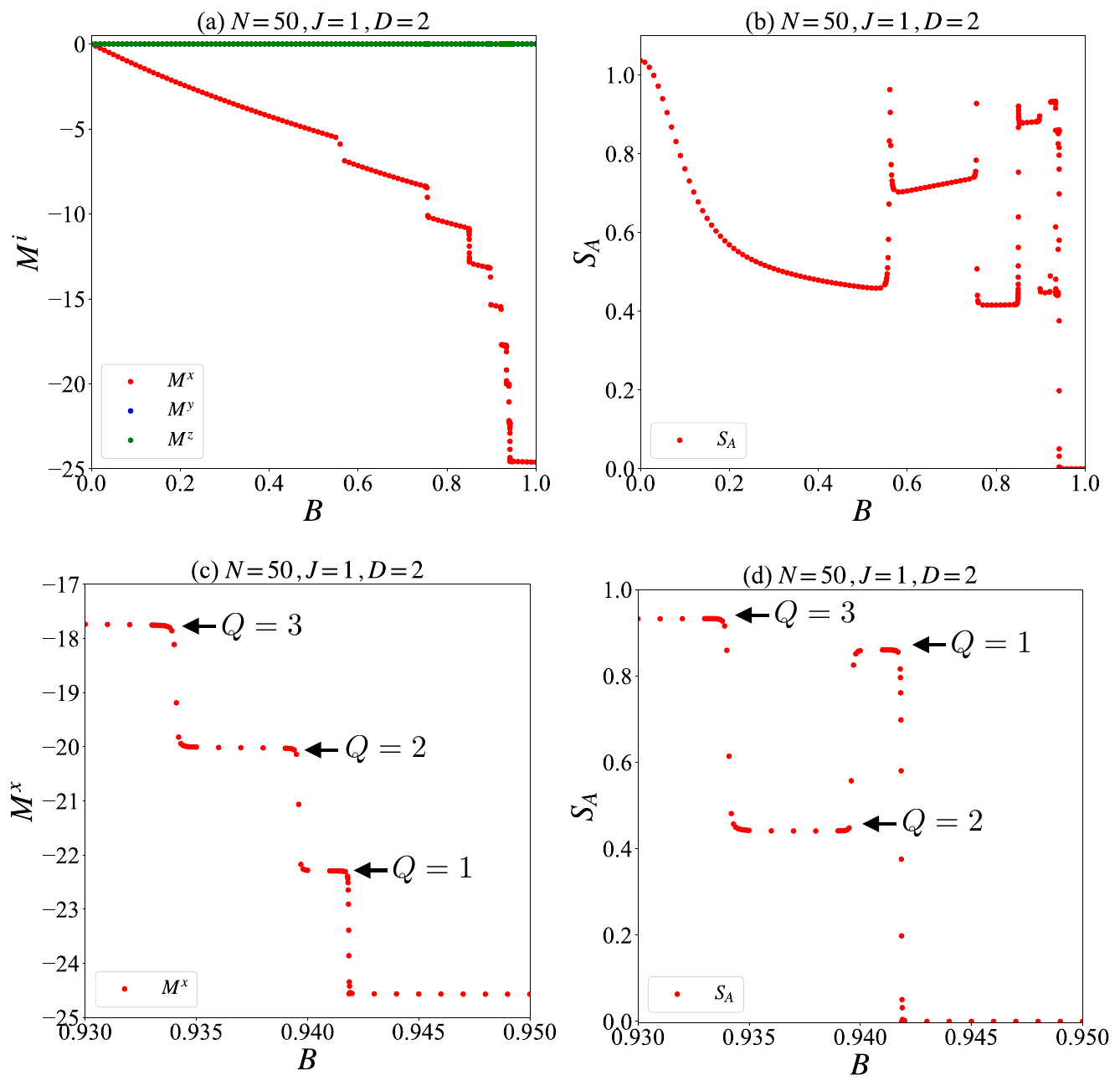}     
    \caption{
    $B$-dependence of the magnetization $M^i$ and entanglement entropy $S_A$. 
    Panel~(a) shows $M^x$, $M^y$, and $M^z$ as functions of $B$. 
    In panel~(b), $S_A$ is plotted with the subsystem size set to $N/2$. 
    Panels~(c) and (d) are magnified views of the panels~(a) and (b), respectively, in the range $0.93 \le B \le 0.95$. 
    The corresponding topological number $Q$ are indicated. 
    }
    \label{fig:Bdep}
\end{figure}

Due to the background inhomogeneity and the transitions between states with different soliton numbers as the magnetic field varies, we find highly nontrivial $B$-dependence of both $M^i$ and $S_A$.
In Fig.~\ref{fig:Bdep} (a), sudden changes are found in the total magnetization which reflect changes in soliton number; however, in the weak-field regime, the soliton overlap suppresses this singular behavior 
(See related studies for discussions of the jump behavior in both classical \cite{PhysRevB.89.014419} and quantum systems \cite{PhysRevA.110.043312,PhysRevB.107.024403,PhysRevB.110.L100403}). 

Figure~\ref{fig:Bdep} (b) shows $S_A$ when the system is bipartitioned at the center.
If a soliton is initially located at the center of the system, a change in the soliton number displaces it from the center.
Since $S_A$ is enhanced where the solitons are present and suppressed in ferromagnetic regions, $S_A$ exhibits a nontrivial $B$-dependence, characterized by alternating large and small values.
Panels~(c) and (d) of Figure~\ref{fig:Bdep}, which present the enlarged views of the region exhibiting sudden change at large $B$, reveal clear plateau structures.
We emphasize that each plateau is characterized by different soliton numbers $Q$ (see Appendix \ref{app:plateau} for details).

\begin{figure}
    \centering
    \includegraphics[width=1.0\linewidth]{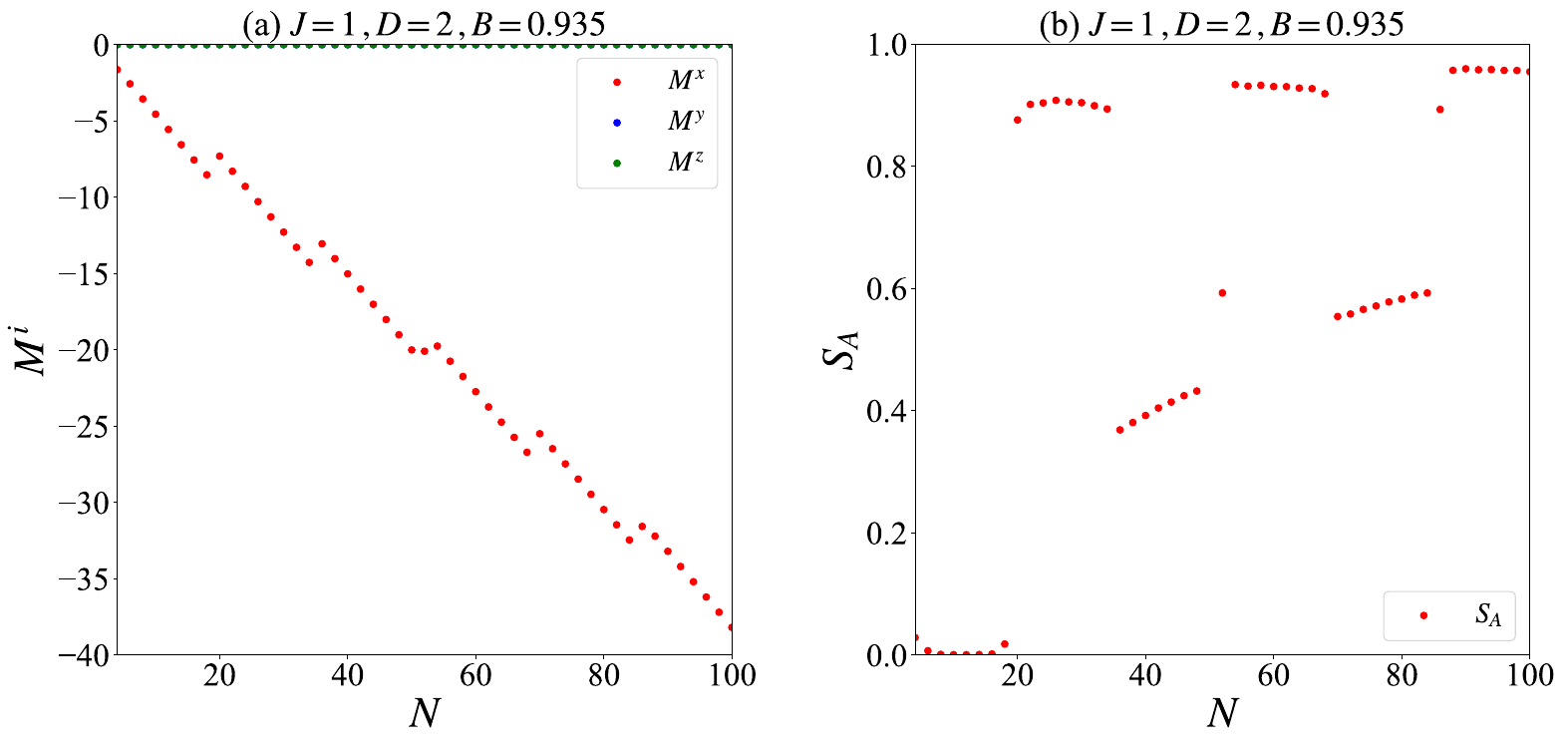}
    \caption{
    Total size dependence of the magnetization $M^i$ and entanglement entropy $S_A$.
    In the panel (a), we plot $M^x$, $M^y$, and $M^z$ as functions of $N$.
    For the plot of $S_A$ in (b), we set the size of the subsystem to be $N/2$.
    }
    \label{fig:SizeDep}
\end{figure}

We also present $N$-dependence of $M^x$ and $S_A$ (Fig.~\ref{fig:SizeDep}). 
As $N$ increases, the soliton emerges into the system since the twisted spin configurations are energetically more favorable than the ferromagnetic alignment.
As a result, the $N$-dependence of $M^x$ and $S_A$ becomes highly nontrivial, especially near the transition points.

As shown in Fig.~\ref{fig:SizeDep}, $S_A$ exhibits the discontinuities that signal the transitions, while the variation in $M^x$ is more gradual near these points.
This suggests that $S_A$ is more sensitive to the transitions than the total magnetization.

These findings link entanglement to topological inhomogeneity. Classically, solitons arise from DM-induced twisting \cite{roessler2006spontaneous}; quantum mechanically, they imprint on $S_A$, suggesting it encodes vacuum structure. This mirrors the sensitivity of the entanglement in perturbed free fields \cite{Casini:2009sr}, but extends to interacting spin systems. The soliton-entropy correspondence holds across $B$, with chirality providing a bridge to classical analogs \cite{braun1996chiral}. Unlike classical phase analysis, which misleads in quantum regimes 
, $S_A$ and $\chi_n$ reliably detect topological features.
Our results imply entanglement entropy can probe chiral magnet ground states, potentially aiding quantum simulation platforms. Future work could explore dynamics or higher dimensions.


\section{Conclusion}
In this study, we explored the entanglement entropy of a monoaxial chiral ferromagnetic quantum spin chain, revealing its sensitivity to the system's topological structure.
Using the DMRG method, we analyzed a spin-1/2 model with exchange interaction, DMI, and Zeeman effect, uncovering spontaneous oscillatory patterns in the entanglement entropy.
These oscillations, tunable by the external magnetic field, correspond closely to the positions of topological solitons, as confirmed by local magnetization and spin chirality profiles.
At low magnetic fields, the entanglement entropy reflects the inhomogeneous background of the chiral soliton lattice, while in high-field ferromagnetic regime, it diminishes to zero, signaling a transition to a trivial, unentangled state. 
The size dependence of the entanglement entropy also shows a nontrivial behavior due to the inhomogeneity. 
Although the maximum value seems to be bounded by $\sim 1$, whether it obeys the area law or not is an open question in this case \cite{RevModPhys.82.277}. 
Our findings establish entanglement entropy as a powerful probe for detecting quantum vacuum inhomogeneity and topological features in chiral magnets, bridging classical soliton physics and quantum correlations.

From the perspective of entanglement entropy, our results suggest that the domain structure of the entanglement is expected to be spontaneously generated in higher dimensions with infinite volume where the Hamiltonian has translational invariance, since our model is known to have the CSL ground state even in such cases. 
The island structures of entanglement are expected to be ubiquitously found in various systems and play crucial roles for understanding the clustering mechanism of quantum information. 
This work not only deepens the understanding of quantum effects in topological systems but also suggests practical applications for characterizing ground states and manipulating entanglement in quantum simulation platforms, opening avenues for future investigations into dynamics and higher-dimensional systems.

\section*{Acknowledgments}
We thank Yusuke Kato for drawing our attention to Refs.~\cite{PhysRevA.110.043312,PhysRevB.107.024403}.
This work is supported by the JSPS Grant-in-Aid for Scientific Research (KAKENHI Grant No.~19K14616, 20H01838, and 25K07156) (RY) and the WPI program ``Sustainability with Knotted Chiral Meta Matter (SKCM$^2$)'' at Hiroshima University (KN and RY).

\appendix

\section{Classical and continuum limit} \label{app:explanation_spin_chirality}

\begin{figure}
    \centering
    \includegraphics[width=0.7\linewidth]{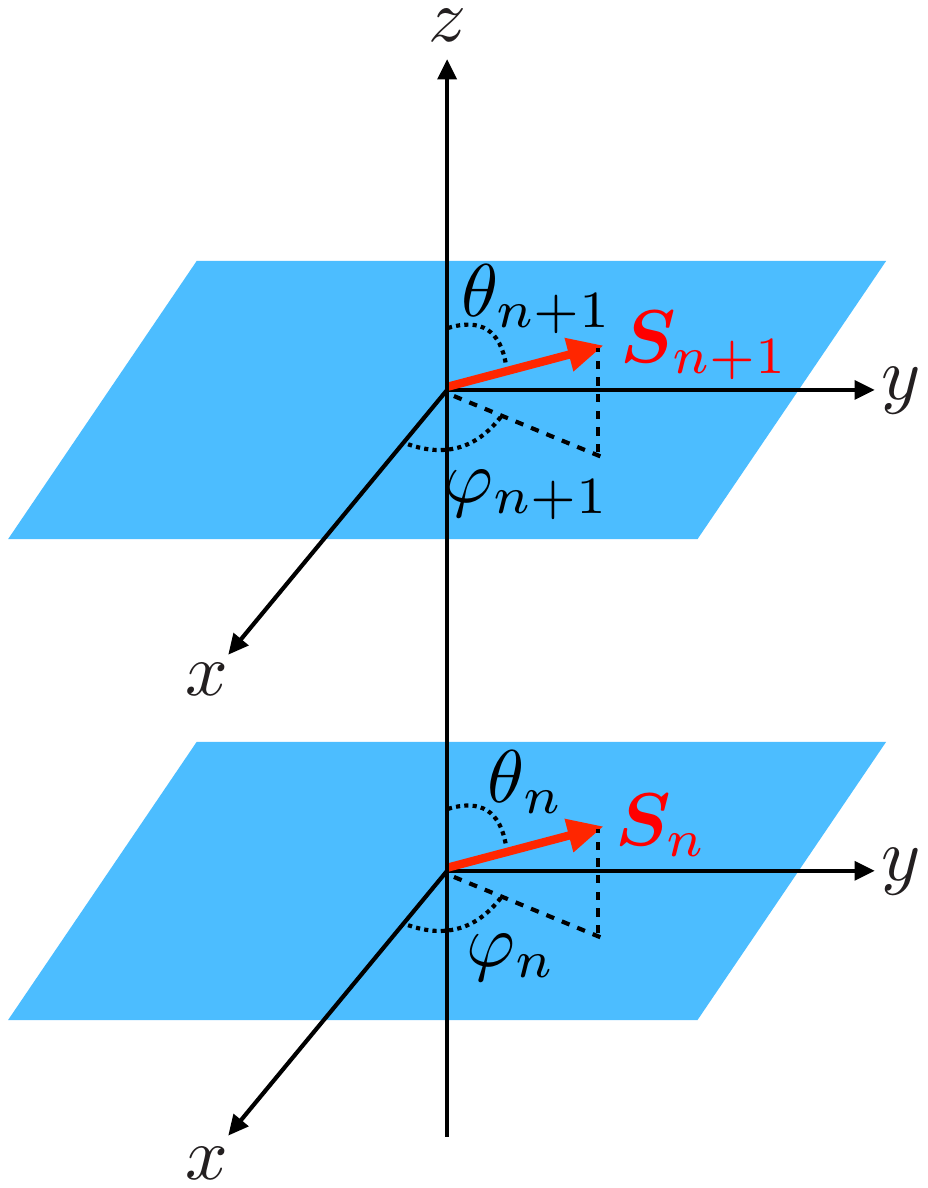}
    \caption{
    Schematic picture of the spin vectors
    }
    \label{fig:schematic_picture_spin_chirality}
\end{figure}

In the main text, we focused exclusively on the $z$-component of the spin chirality. 
In this appendix, we clarify the meaning of the $x$- and $y$-components and confirm that they are not relevant in the context of our study.
A schematic illustration of the spin vector in polar coordinates is shown in Fig.~\ref{fig:schematic_picture_spin_chirality}.
We assume $\varphi_n = \pi/2$ for all $n$ to simplify the discussion. 
In this case, the spins lie in the $yz$-plane, and only the $x$-component of the spin chirality becomes nonzero, given by $\chi^x_n = S^2\sin(\theta_n - \theta_{n+1})$. 
Thus, $\chi^x_n$ quantifies the rotation of spin vectors within the $yz$-plane. 
Analogously, $\chi^y_n$ reflects the rotation in the $zx$-plane.

\begin{figure}
    \centering
    \includegraphics[width=1.0\linewidth]{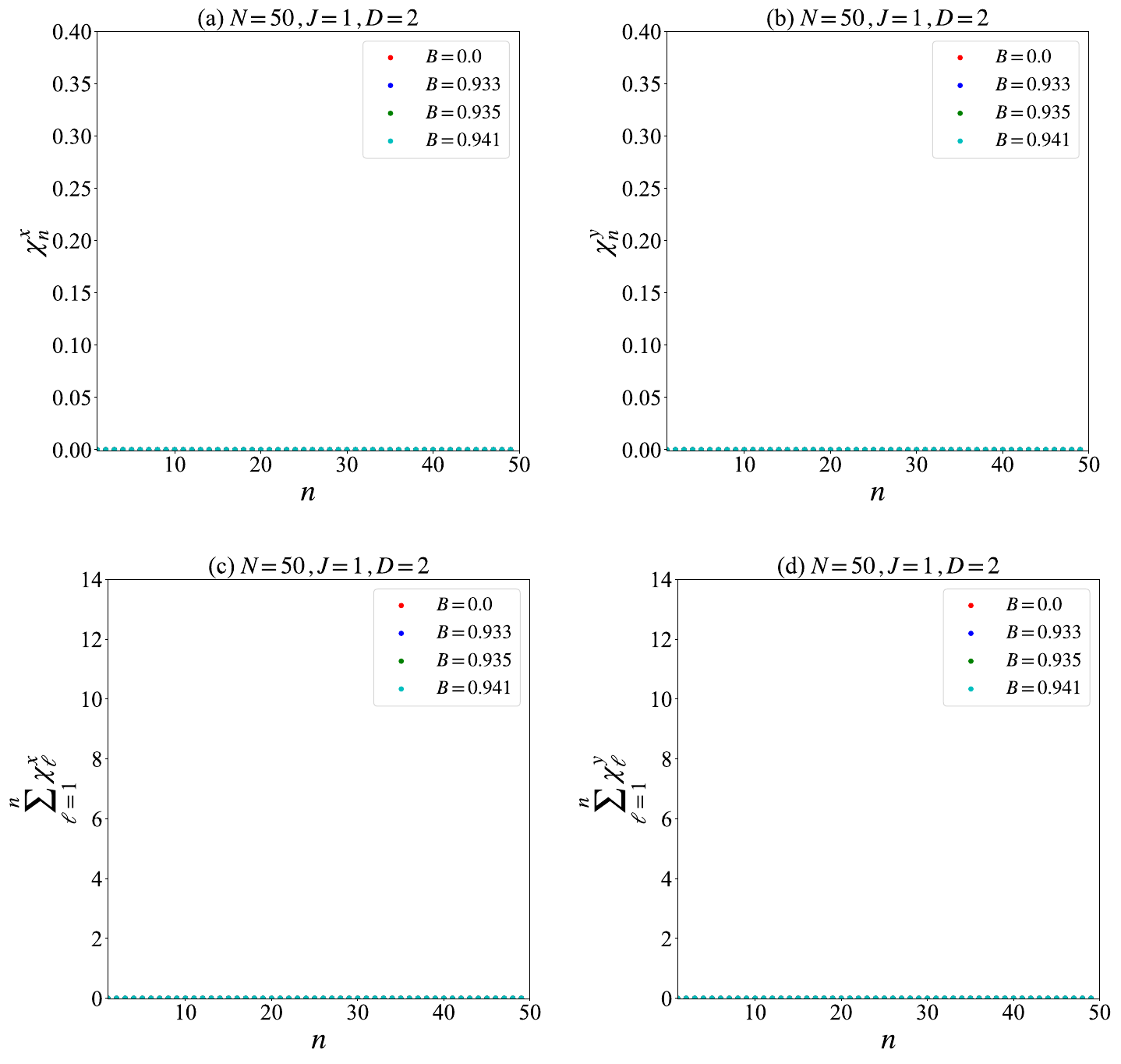}
    \caption{
    Plots of $\chi^x_n$ (a) and $\chi^y_n$ (a) at site $n$ and the accumulated spin chiralities $\sum_{\ell=1}^n\chi^x_\ell$ (c) and $\sum_{\ell=1}^n\chi^y_\ell$ (d). 
    }
    \label{fig:spin_chirality_xy}
\end{figure}

Figure~\ref{fig:spin_chirality_xy} shows the numerical results for $\chi^x_n$ and $\chi^y_n$, indicating that both components remain approximately zero and show no dependence on $n$, indicating that the spin expectation values do not rotate in the $yz$- or $zx$-planes. 
This justifies our choice to focus only on $\chi^z_n$ in the main text.

We now discuss the continuum and classical limits of $\chi^z_n$. 
In the classical limit, the spin operator $\hat{S}^\mu_n$ is replaced by a classical spin vector $(S^x_n, S^y_n,S^z_n)= S \bm{n}_n$ with $S = 1/2$, where the unit vector $\bm{n}_n$ is parametrized in polar coordinates as 
$
\bm{n}_n = (\sin \theta_n \cos \varphi_n,\, \sin \theta_n \sin \varphi_n,\, \cos \theta_n)^\mathsf{T}
$.

We adopt the spin chirality to characterize solitonic structures rather than employing the soliton number operator introduced in Ref.~\cite{PhysRevB.107.024403}. 
While this operator is conserved when $J = 0$ and is helpful in analytically constructing eigenstates, it is not conserved in our case with $J \neq 0$.
Therefore, we focus on the spin chirality, which is related to the topological charge of the classical sine-Gordon solitons, as discussed below.
Taking the continuum limit (with lattice spacing $a \to 0$ and $z = na$), the spin chirality becomes
\begin{gather}
    \chi^z(z) = S^2 \sin^2 \theta(z) \frac{\rmd \varphi(z)}{\rmd z} \,.
\end{gather}
In particular, when the spins lie in the $xy$-plane ($\theta = \pi/2$), $\chi^z(z)$ simply represents the spatial variation of the phase $\varphi(z)$.
Thus, large values of $\chi^z(z)$ correspond to regions where the spins are strongly twisted—i.e., the locations of the sine-Gordon solitons.

Finally, we emphasize that the quantity
\begin{gather}
    N_{\mathrm{topo}} := \frac{1}{2\pi} \int \rmd z \frac{\rmd \varphi(z)}{\rmd z} \in \mathbb{Z}
\end{gather}
is the topological charge, provided that $\varphi(z)$ takes the same value (mod $2\pi$) at both boundaries.
Naturally, the quantized topological number discussed above is well-defined only in the classical and continuum limits. 
Nonetheless, a key advantage of the spin chirality is that it offers an intuitive understanding of such structures within these limiting regimes.

\section{Numerical results of local magnetization} \label{app:local_magnetization}

\begin{figure}
    \centering
    \includegraphics[width=1\linewidth]{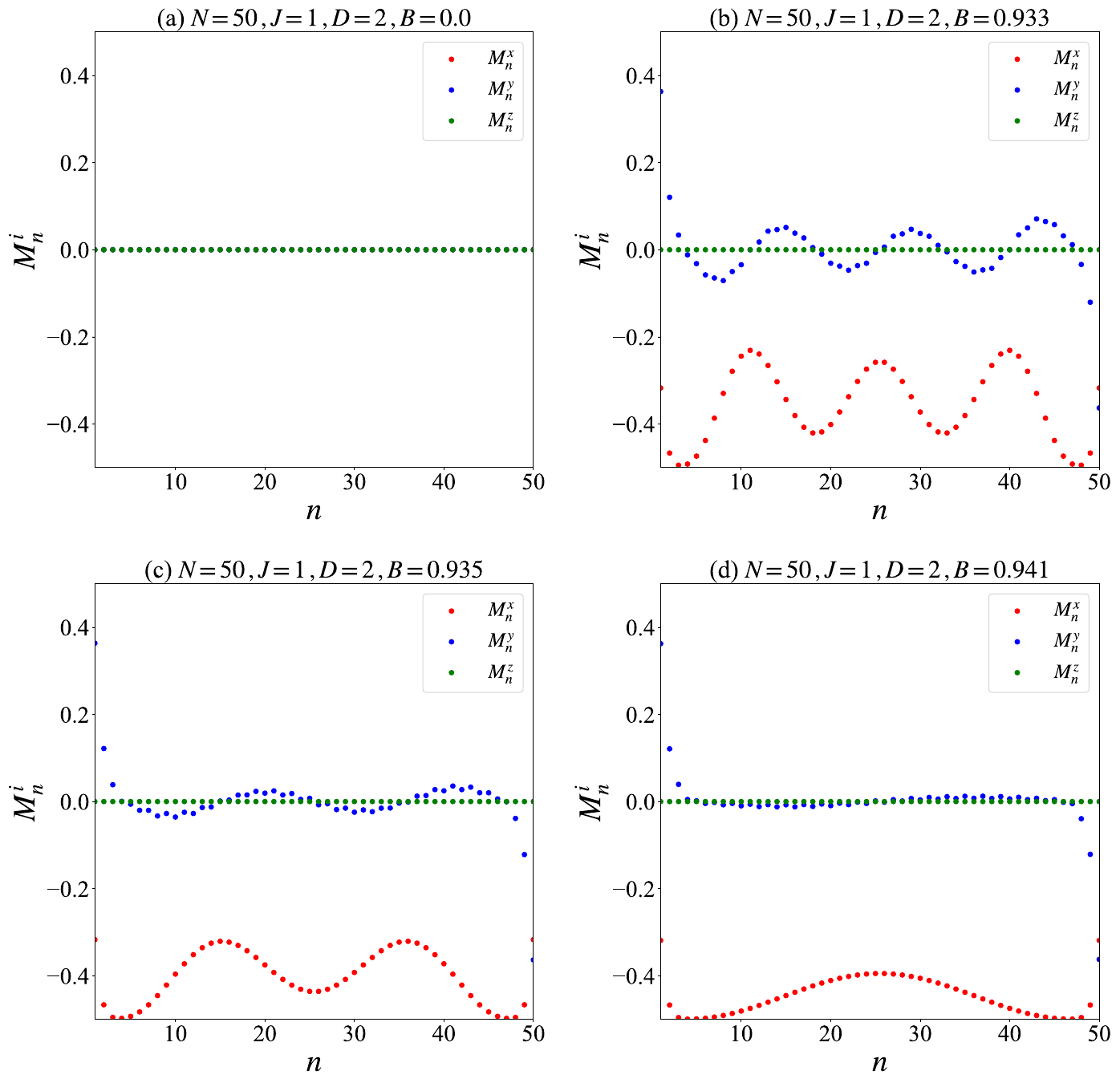}
    \caption{
    The spatial dependencies of the magnetization $M^i_n$ are shown in the panels for $B=0.0$ (a), $B=0.933$ (b), $B=0.935$ (c), and $B=0.941$ (d).
    }
    \label{fig:magnetization}
\end{figure}

In this appendix, we present numerical results for the spatial dependence of the magnetization $M^i_n$. 
Panel~(b) of Fig.~\ref{fig:magnetization} shows that for $B = 0$, $M^i_n$ vanishes and is independent of $n$, indicating a degeneracy in the spin expectation values. 
In the classical counterpart, the spin configuration forms a helimagnetic structure, where the spins rotate around the $z$-axis with a constant pitch angle. 

As $B$ increases, the spins tend to align along the $-x$ direction due to the Zeeman term, leading to an increase in $|M^x_n|$. Panels~(b), (c), and (d) of Fig.~\ref{fig:magnetization} confirm that $|M^x_n|$ increases and displays spatial oscillations, while $M^y_n$ oscillates around zero.
The number of maxima in $M^x_n$ is 3, 2, and 1 for $B = 0.933$, $0.935$, and $0.941$, respectively.
For sufficiently large $B$, except near the boundaries, the spins align along the $-x$ direction, in agreement with classical expectations.

\section{Plateau structure in $B$-dependence of $M^x$ and $S_A$}
\label{app:plateau}

\begin{figure}
    \centering
    \includegraphics[width=1.0\linewidth]{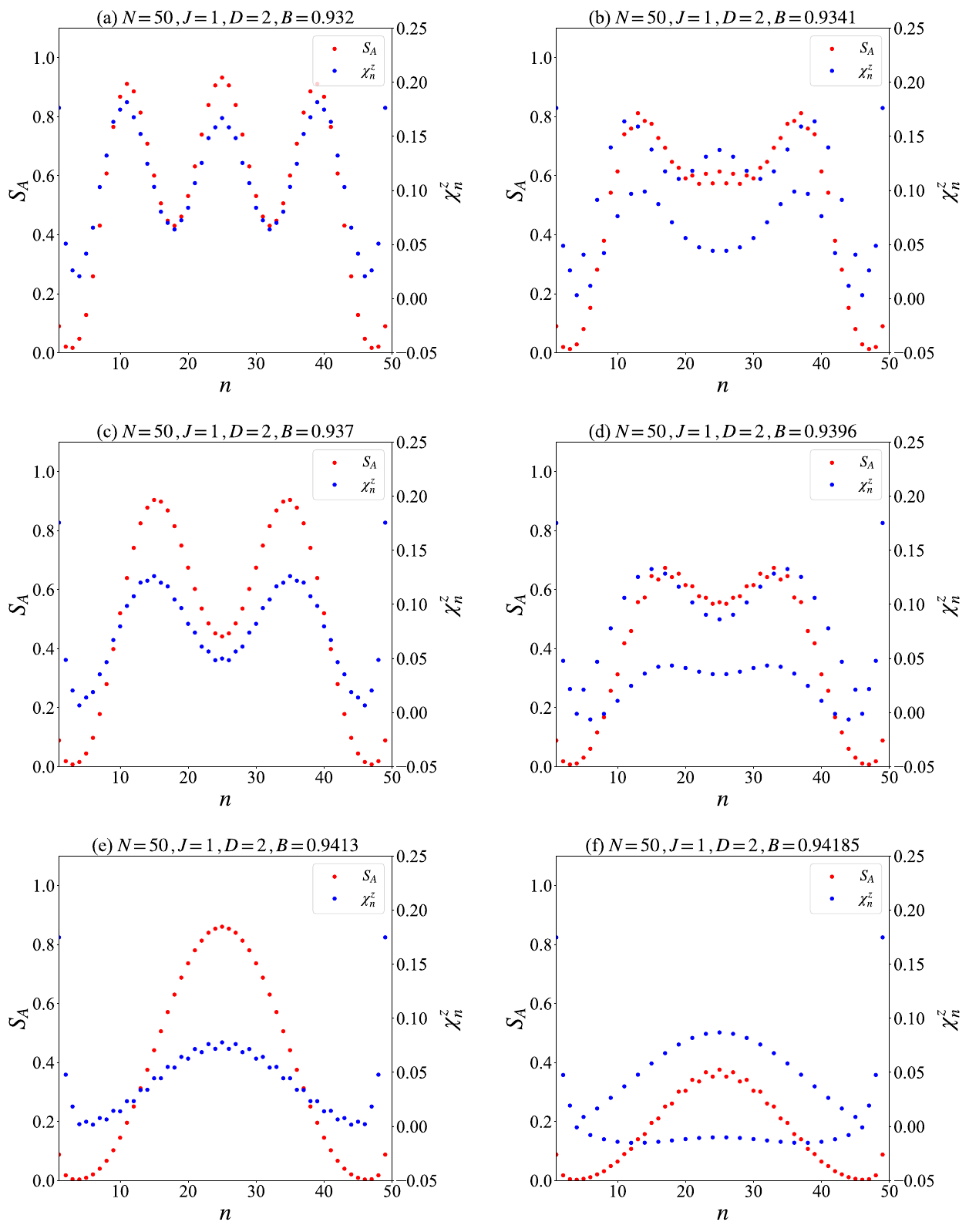}
    \caption{
    Plots of the entanglement entropy $S_A$ and $z$-component of the spin chirality $\chi^z_n$ for $B=0.932$ (a), $B=0.9341$ (b), $B=0.937$ (c), $B=0.9396$ (d), $B=0.9413$ (e), and $B=0.94185$ (f).
    }
    \label{fig:plateau_detail}
\end{figure}

In the main text, we presented $\chi^z_n$ and $S_A$ for a representative point within each soliton-number plateau labeled by $Q$.
To further confirm these results, we now show the $B$-dependence of $\chi^z_n$ and $S_A$ for additional values of $B$, as shown in panels (a), (c), and (e) of Figure~\ref{fig:plateau_detail}.
For magnetic fields corresponding to the soliton number $Q$, $\chi^z_n$ exhibits $Q$ local maxima.
However, there is some subtlety in defining the soliton number $Q$.

To demonstrate this, we examine $\chi^z_n$ and $S_A$ at intermediate values of $B$ where the system undergoes a transition between different soliton numbers. 
For instance, panel~(b) of Figure~\ref{fig:plateau_detail} corresponds to the transition from $Q = 3$ to $Q = 2$.
In this regime, $\chi^z_n$ exhibits strong oscillations as a function of $n$, while $S_A$ takes smaller values than in panel~(a) and displays fine fluctuations near the center of the system.
These fine oscillations make it difficult to count the number of maxima unambiguously.

Similar behavior is also observed in panels~(d) and (f), corresponding to the transitions from $Q = 2$ to $Q = 1$ and from $Q = 1$ to $Q = 0$, respectively.
In fact, the large oscillations of $\chi^z_n$ can be attributed to the fact that $\chi^z_n$ takes negative values in panel~(f).
Such negative values indicate that the spin twist is being unwound.
As a result, the ground state tends toward the ferromagnetic configuration induced by the magnetic field, and the entanglement entropy $S_A$ is significantly reduced compared to that in panel~(e).

\bibliography{reference.bib}


\end{document}